\documentclass[amsmath,amssymb,aps,prd,11pt,tightenlines,superscriptaddress, 
nofootinbib,preprintnumbers,notitlepage]{revtex4-1}

\usepackage{graphicx}
\usepackage{subcaption}
\usepackage{float}
\usepackage{dcolumn}
\usepackage{bm}
\usepackage{amssymb}
\usepackage{amsmath}
\usepackage{mathrsfs}
\usepackage{graphicx,color} 
\usepackage[english]{babel}     \usepackage{float}
\usepackage{graphicx,wrapfig,lipsum}      
\usepackage{multimedia}    
\usepackage[mathscr]{eucal}  
\usepackage{bm}              
\usepackage{amssymb}           
\usepackage{amsmath}      
\usepackage{mathrsfs}  
\usepackage[mathscr]{eucal}      
\usepackage[normalem]{ulem}
\usepackage{eucal}

\begin{document}

\title{Thick oriented and nonoriented center-vortex $SU(N)$ configurations with fractional topological charge lumps}
\author{David R. Junior}
\affiliation{
Instituto de F\'\i sica, Universidade Federal Fluminense, Niter\'oi, 24210-346, RJ, Brazil} 
\affiliation{
Instituto de F\'isica Te\'orica, Universidade Estadual Paulista and South-American Institute of Fundamental Research ICTP-SAIFR, Rua Dr. Bento Teobaldo Ferraz, 271 - Bloco II, S\~ao Paulo, 01140-070, SP, Brazil}
\affiliation{Institut f\"ur Theoretische Physik, Universit\"at T\"ubingen, Auf der Morgenstelle 14, T\"ubingen, 72076,  Germany}
\author{Luis E. Oxman}
\affiliation{
Instituto de F\'\i sica, Universidade Federal Fluminense, Niter\'oi, 24210-346, RJ, Brazil} 
\author{Gustavo M. Sim\~oes}

\affiliation{Instituto de Física de São Carlos, Universidade de São Paulo, IFSC–USP,
 São Carlos, 13566-590, SP, Brazil}
\date{\today} 
\begin{abstract}

Mixed oriented and nonoriented center vortices are known to generate nontrivial topological charge. However, most previous analyses have been restricted to Abelian-projected thin configurations. Studies of thick vortices have so far focused on the $SU(2)$ case and on the intersection of a single pair of oriented objects. In this work, we construct mixed oriented and nonoriented thick center-vortex gauge fields in $SU(N)$ with smooth profiles and explicit non-Abelian phases. These phases ensure a smooth interpolation between different Cartan fluxes at monopole junctions. We analyze the resulting topological charge density and visualize its morphology, elucidating the color structures responsible for fractional lumps that together yield a nonvanishing global charge.
  
\end{abstract}
\maketitle
\section{Introduction}

The vacuum structure of Yang–Mills (YM) theory is shaped by nontrivial topological field configurations, which play a central role in its nonperturbative dynamics. The distribution of topological charge density has been visualized in \cite{Ilgenfritz:2007xu}, providing direct insight into the underlying gauge field fluctuations. A key quantity in this context is the YM topological susceptibility obtained as the quantum average:
\begin{gather}
\chi = \int d^4x\, \langle q(x) q(0)\rangle  =\langle Q^2\rangle/V \;,\nonumber \\
Q= \int d^4x\, q(x) \makebox[.5in]{,}   q(x) = \frac{1}{32\pi^2 N}(F_{\mu\nu},\tilde{F}_{\mu\nu}) \;,
    \label{tc}
\end{gather}
 measured in Ref. \cite{DelDebbio:2002xa}. 
Here, $q(x)$ is the topological charge density and $V$ is the Euclidean spacetime volume\footnote{The Killing product $(X,Y)$ between two Lie Algebra elements is defined with the   convention $(T_A,T_B)=\delta_{AB}$.}.

Configurations with nontrivial charge have long been regarded as relevant for understanding nonperturbative phenomena. Indeed, by the Atiyah–Singer index theorem, $Q$ is related to the difference between the number of left- and right-handed zero modes of the Euclidean Dirac operator. In turn, the spectral eigenvalue density of the Dirac operator near zero determines the chiral condensate through the Banks–Casher relation~\cite{BanksCasher}. Thus, fluctuations in the topological sector of the gauge field ensemble control the distribution of low-lying Dirac eigenvalues and thereby are expected to drive chiral symmetry breaking in the QCD vacuum.

Among the gauge field configurations that contribute to the topological charge, center vortices are the most prominent, as they are also strongly believed to play a fundamental role in the confinement mechanism of pure YM theory \cite{Greensite:2003bk}. From the lattice perspective, center-projection methods have been employed to isolate the vortex content of YM configurations. Removing center vortices from gauge configurations suppresses confinement~\cite{DelDebbio:1997ke}, the topological susceptibility, as well as chiral symmetry breaking~\cite{vortex-removed}. However, globally oriented center vortices have vanishing $Q$, so nonoriented configurations are essential for generating a nontrivial topological charge and, hence, a nonvanishing susceptibility~\cite{top-2,vortex-en}. Indeed, in these works, the topological charge density is shown to arise at intersection points between thin center-vortex sheets or from writhing segments. In particular, at the intersections, the topological charge in $SU(3)$ was found to take the values $1/3$ and $2/3$ (see also Refs. \cite{C-1,H-2} and \cite{C-2}). The origin of the writhe contribution was clarified in Ref. \cite{top-3} through regularized calculations in the $SU(2)$ case. Remarkably, after appropriate smoothing to account for nonorientability, vortex-only ensembles in $SU(2)$ Yang--Mills theory were shown to reproduce the full nonperturbative topological susceptibility~\cite{top-1}. The nonoriented component, where center vortices in the network contain non-Abelian monopoles with adjoint Cartan flux, was also shown to be essential to drive the formation of the confining flux tube between quark probes in pure $SU(N)$ YM theory \cite{Oxman:2021yij}. Recently, nonoriented center vortices were shown to be preferred saddle points when the theory is defined in compactified spacetimes \cite{yuya}. In this context, fractional topological excitations in the $CP^{N-1}$  model on a two-dimensional torus with nontrivial ’t Hooft twist have been explicitly constructed and analyzed in Ref. \cite{compact-2}, where their relation to twisted boundary conditions and fractional sectors is discussed.

Center vortices are intrinsically thick objects, treating them as thin is merely a consequence of the center projection used in the detection method. The topological charge of thick configurations has been studied for $SU(2)$~\cite{top-2}, in the case of a single self-intersecting oriented vortex, where the thickness is accounted for at the level of the field strength.  To date, there are no calculations of the topological charge for thick mixed oriented and nonoriented configurations starting from the underlying non-Abelian gauge field configurations. In addition, regarding charge densities,  lattice $SU(3)$ simulations  displayed lava-lump fluctuations \cite{visual}. In Ref. \cite{frac-lump}, these lumps were shown to mainly have $Q=1/3$ and possible theoretical frameworks involving instanton-dyon configurations were discussed. Obtaining typical densities in mixed thick configurations requires a careful treatment of the non-Abelian phases that characterize them. Here, we shall explicitly construct these phases for mixed oriented and nonoriented center-vortex configurations in $SU(N)$. This will allow for an appropriate smoothing procedure to generate thick objects and study the corresponding topological charge densities. These steps are in the spirit of the Cho–Faddeev–Niemi decomposition \cite{Manton}–\cite{Shaba}, extended to include collimated objects. To this end, besides monopole-like defects, the local color frame contains defects at the center-vortex guiding centers \cite{conf-qg}. In addition, these procedures will allow us to understand the observed topological charge carried by lattice density lumps in light of effective models for center-vortex ensembles.  

In Sec. \ref{rev}, we review the main differences between oriented and nonoriented center vortices. Next, in Sec. \ref{mixed-thin-S}, we discuss the topological charge of non-Abelian $SU(N)$ gauge fields for thin mixed oriented and nonoriented center vortices. Sec.~\ref{mixed-thick} is devoted to the study of the topological charge density of mixed configurations, considering thick objects. These collimated finite-size fluxes are closer to configurations expected to dominate the YM vacuum. Finally, in Sec.~\ref{sec:concl}, we present our conclusions.

\section{Collimated center-vortex configurations}
\label{rev}

Collimated flux configurations involving center vortices and Cartan monopoles on them can be written in the general form 
\cite{conf-qg}
\begin{gather}
    \mathrm{Ad} (A_\mu) = R \mathrm{Ad}(\mathcal{A}_\mu) R^{-1}+iR\partial_\mu R^{-1}\label{Ad-g} \;,
\end{gather}
where $\mathrm{Ad}(\cdot)$ refers to the adjoint representation of $SU(N)$. In particular, 
$R=\mathrm{Ad}(S)$ can be associated with a local frame in the Lie algebra,
\begin{gather}
  n_A =  ST_A S^{-1} = T_B R_{BA}(S) \;,
\end{gather}
that contains defects but is single-valued when going around any loop. On the other hand, when following a loop that links the center-vortex worldsurface, $S$ changes by a center-element $e^{\pm i\frac{2\pi k}{N}}$, $k=0, \dots, N-1$. For thin objects $\mathcal{A}_\mu =0$, while for thick configurations $\mathcal{A}_\mu $ must be such that $A_\mu$ and the field-strength are smooth. 

In order to write well-defined expressions in the defining representation, it is convenient to introduce the frame-dependent fields $C_\mu^A$, $A= 1, \dots, N^2-1$, through
\begin{align}
\mathrm{Ad} (C_\mu)= C_\mu^A M_A =i R^{-1}\partial_\mu R \makebox[.5in]{,}
M_A = \mathrm{Ad} (T_A)  \;.
\end{align}
In this manner,
\begin{gather}
    A_\mu = (\mathcal{A}_\mu^A -C_\mu^A) n_A \;,
\end{gather}
and the associated field-strength is
\begin{align}
F_{\mu \nu}=G_{\mu \nu}^A n_A   \makebox[.3in]{,}   G^A_{\mu\nu}=F_{\mu \nu}^A(\mathcal{A})- F^A_{\mu \nu}(C)\;\;.
    \label{Fst}  
\end{align}
 Let us quote two possible $S$-mappings, those corresponding to isolated oriented or nonoriented configurations. In both cases, the center-vortex guiding-centers are associated with defects in the local off-diagonal directions $n_\alpha$, $n_{\bar{\alpha}}$ while, in the nonoriented case, the monopole locations are given by defects in the local Cartan directions $n_q$:
\begin{equation}
n_q = S T_q S^{-1} 
\makebox[.5in]{,}
n_{\alpha} = S T_{\alpha} S^{-1}
\makebox[.5in]{,}
n_{\bar{\alpha}} = S T_{\bar{\alpha}} S^{-1}\;.
\label{Tqn}
\end{equation}
Here, $T_q$, $q=1, \dots, N-1$ are the Cartan generators of $\mathfrak{su}(N)$ and $T_\alpha = (E_\alpha + E_\alpha^\dagger)/\sqrt{2}$, $T_{\bar{\alpha}}= -i(E_\alpha - E_\alpha^\dagger)/\sqrt{2}$ are defined in terms of the root vectors,
 \begin{gather}
[T_q, E_\alpha]= \alpha|_q\, E_\alpha \;.
\end{gather}
 
 A single oriented vortex with guiding center on $\tilde{\Sigma}$ can be generated with the Cartan mapping 
\begin{align}
    \tilde{C} = e^{i\chi \beta\cdot T} \makebox[.5in]{,} \beta \cdot T = \beta|_q T_q\;,
    \label{oriented}
\end{align}
where $\chi$ changes by $2\pi$ when following a loop $\mathcal{C}$ that links $\tilde{\Sigma}$. For elementary center vortices, the $(N-1)$-tuple $\beta$ is one of the magnetic weights of the defining representation $\beta_1, \dots , \beta_N$. The gauge field for a thin vortex reads
\begin{align}
    A =\partial_\mu\chi \,\beta\cdot T\label{thin-ori}\;.
\end{align}
For elementary center vortices, the  Wilson loop along $\mathcal{C}$ yields
\begin{align}
    W[A]=z^{L(\mathcal{C},\tilde{\Sigma})}\makebox[.5in]{,}z=e^{i\frac{2\pi \beta\cdot T}{N}}=e^{-i\frac{2\pi}{N
    }}I\;,
    \label{link}
\end{align}
where $L(\mathcal{C},\tilde{\Sigma})$ is the linking number between $\mathcal{C}$ and $\tilde{\Sigma}$. The field-strength of this vortex reads
\begin{align}
    F_{\mu\nu}=2\pi \beta\cdot T\int_{\tilde{\Sigma}} d\tilde{\sigma}_{\mu\nu}\, \delta^{(4)}(x-\bar{x})\;,
\end{align}
and is thus localized on the closed surface $\tilde{\Sigma}$. 
 In this case, the local Cartan directions are in fact global $n_q =T_q$, while the off-diagonal ones
\begin{gather}
    n_\alpha = \tilde{C} T_\alpha \tilde{C}^{-1} \makebox[.5in]{,} n_{\bar{\alpha}} = \tilde{C} T_{\bar{\alpha}} \tilde{C}^{-1} \;,
    \end{gather}
have a defect at $\tilde{\Sigma}$. More precisely, for a given defining magnetic weight $\beta$, some of the roots $\alpha$ satisfy $\beta \cdot \alpha = \pm 1$, that is,
    \begin{gather}
[\beta \cdot T ,  E_\alpha]= \pm  E_\alpha \;.
\end{gather}
Then, $n_\alpha$, $n_{\bar{\alpha}}$ rotate once when going around $\Sigma$. The remaining root vectors commute with $\beta \cdot T$ and the corresponding frame directions are constant.

A typical nonoriented configuration is characterized by the non-Abelian $S$-mapping \cite{conf-qg}, 
\begin{gather}
    S_{\rm n}= C W\makebox[.5in]{,}
   C= e^{i\varphi \beta \cdot T}
\makebox[.5in]{,}
W = e^{i\theta \sqrt{N} T_{\alpha} }  \;,
\label{SW}
\end{gather}
where $\varphi$ and $\theta$  are spherical angles.
Around the north pole ($\theta \sim 0$), 
\[
S_{\rm n} \sim  e^{i\varphi \beta \cdot T} \;,
\]
while close to the south pole, $W(\theta)\sim W(\pi)$ is a Weyl reflection. For $\alpha = (\beta -\beta')/2N $:
\begin{gather}
W(\pi)^{-1} \beta \cdot T W(\pi)=\beta' \cdot T
\makebox[.5in]{,}
\beta' = \beta-2\alpha (\alpha \cdot \beta)/\alpha^2 \makebox[.5in]{,}    S\sim W(\pi)\, e^{i\varphi\, \beta' \cdot T}  \;.
\label{Weyl-t}
\end{gather}
Because of these properties, different off-diagonal components of the local frame are rotated close to the north and south pole, namely, those characterized by $\beta \cdot \alpha = \pm 1$ or  $\beta' \cdot \alpha' = \pm 1$, respectively. This is reflected in the change of orientation of the projected field-strength components along $n_q$ and in the charge of the interpolating monopole
\[
Q_{\rm m} = 2\pi\, ( \beta   -\beta'  ) = 2\pi 2N \alpha \;.
\]
The contribution of nonoriented configurations to the Wilson loop is also given by Eq. \eqref{link}, as it only depends on the center-element generated by the Cartan factor in Eq. \eqref{SW}. Note also that acting on the left of $S$  with a regular $U$, namely $S\to US$, leads to a physically equivalent configuration. The topological-charge density remains unchanged under this gauge transformation.

\section{Mixed thin oriented and nonoriented center vortices}
\label{mixed-thin-S}

As discussed in \cite{top-2}, mixed oriented and nonoriented vortex configurations generate a nontrivial fractional topological charge. However, no explicit expression for the associated gauge field was given. Here, we shall consider the map 
\begin{align}
    S=   S_{\mathrm{n}} \tilde{C} \;, 
\end{align}
where the first factor corresponds to the nonoriented center vortex in Eq. \eqref{SW} and the second factor, defined in Eq. \eqref{oriented}, gives the oriented part. The nonconstant Lie Algebra structure is manifested in the associated thin configurations
\begin{gather}
A_\mu = \partial_\mu \chi\,  S_{\mathrm{n}} (\beta\cdot T) S_{\mathrm{n}}^{-1} +\partial_\mu \varphi\, \beta\cdot T +\partial_\mu  \theta\, \sqrt{N}C T_\alpha C^\dagger \;,\label{nonori} \\
F_{\mu\nu}=2\pi S_{\mathrm{n}} (\beta\cdot T)S_{\mathrm{n}}^{-1} \int_{\tilde{\Sigma}}d\tilde{\sigma}_{\mu\nu}\, \delta^{(4)}(x-\bar{x}) +2\pi\beta\cdot T \int_{\Sigma} d\tilde{\sigma}_{\mu\nu}\, \delta^{(4)}(x-\bar{x})\;. \label{nonori-1}
    \end{gather} 

\subsection{Topological charge}

Suppose the nonoriented vortex is placed on the surface $\Sigma$ defined by the $z$-axis for every Euclidean time. In addition, for a simpler visualization, $\tilde{\Sigma}$ is contained in the $\mathbb{R}^3$ time-slice at $t=0$. 
From Eq. \eqref{nonori-1}, the topological charge of the thin mixed configuration is:
\begin{gather}
Q =\big( I(\Sigma,\Sigma) + I(\tilde{\Sigma},\tilde{\Sigma}) +I(\Sigma, \tilde{\Sigma}^+)\big)\frac{\beta^2}{2N}
    + \frac{I(\Sigma, \tilde{\Sigma}^-)}{2N} \big(W^\dagger(\pi)(\beta\cdot T)W(\pi) , \beta \cdot T) \big)\;.
\end{gather}
where $I(\Sigma, \tilde{\Sigma})$ is the intersection number between $\Sigma$ and $\tilde{\Sigma}$,
\begin{align}
    I(\Sigma, \tilde{\Sigma}) = I(\tilde{\Sigma}, \Sigma) = \frac{1}{2}\int_{\Sigma} d^2\sigma_{\mu\nu}\int_{\tilde{\Sigma}} d^2\tilde{\sigma}_{\mu\nu}\delta^{(4)}(x(s)-x(s'))\;,
\end{align}
and we have denoted $\tilde{\Sigma}^+$ and $\tilde{\Sigma}^-$ as the parts of $\tilde{\Sigma}$ that lie in the $z >0$ and $z<0$ regions, respectively. Now, using Eq. \eqref{Weyl-t} and that for any oriented closed surface $\Sigma$, when properly regularized, the self-intersection $I(\Sigma,\Sigma)$ vanishes, the topological charge is
\begin{align}
   Q&=\frac{\beta^2}{2N}I(\Sigma, \tilde{\Sigma}^+)+\frac{\beta\cdot\beta'}{2N} I(\Sigma, \tilde{\Sigma}^-) =\frac{N-1}{N} I(\Sigma, \tilde{\Sigma}^+)-\frac{1}{N}I(\Sigma, \tilde{\Sigma}^-)\;.\label{thin-top-charge}
\end{align}

\begin{figure}[htbp]
    \centering
    \begin{subfigure}[b]{0.45\textwidth}
        \centering
        \includegraphics[scale=0.3]{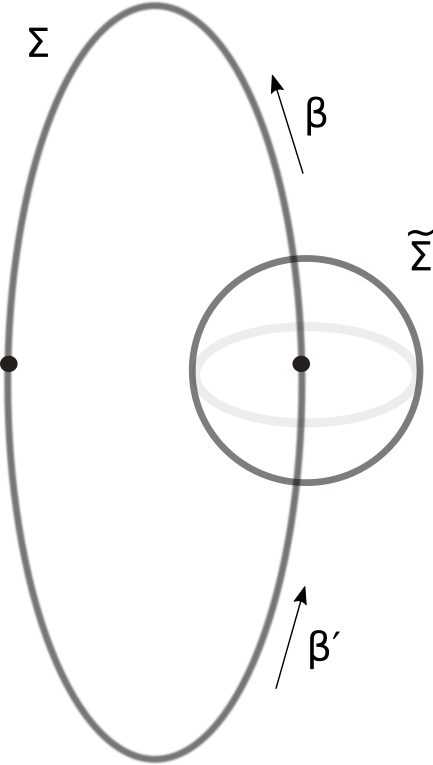}
        \caption{Thin vortices: the monopole pair maps a loop along the time-direction (not displayed). }
        \label{mixed-thin}
    \end{subfigure}
    \hfill
    \begin{subfigure}[b]{0.45\textwidth}
        \centering
        \includegraphics[scale=0.4]{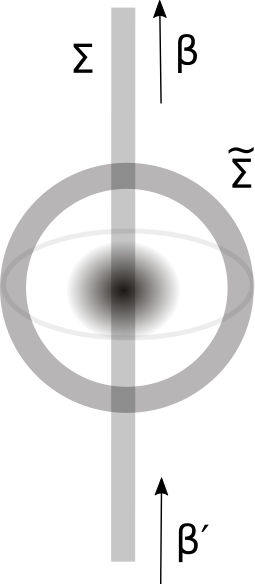}
        \caption{ Thick vortices: the nonoriented vortex is on the $z$-axis for every Euclidean time.}
        \label{mixed}
    \end{subfigure}
    \caption{Mixed oriented and nonoriented center
    vortices displayed at a fixed $t=0$ time-slice.}
    \label{fig:non}
\end{figure}

In Ref. \cite{C-2}, a nonzero topological charge was generated from a configuration containing a nexus and a center-vortex. In particular, the fundamental nexus for $SU(2)$ was defined from the singular map $V=e^{i \varphi\frac{ \sigma_i}{2} \hat{r}_i}$.  The covariant dual field strength of the nexus is given by
\begin{align}
    &-\tilde{F}_{\mu\nu}= \pi \sigma_3 \int_{\Sigma_1} d^2\tau_{\mu\nu}\delta^{(4)}(x-y(\tau_1,\tau_2))-\pi \sigma_3 \int_{\Sigma_2} d^2\tau_{\mu\nu}\delta^{(4)}(x-y(\tau_1,\tau_2))\;,\nonumber\\&
    d^2\tau_{\mu\nu}=d\tau_1d\tau_2\left(\frac{\partial y_\mu}{\partial \tau_1}\frac{\partial y_\nu}{\partial \tau_2}-\frac{\partial y_\mu}{\partial \tau_2}\frac{\partial y_\nu}{\partial \tau_1}\right)\;.
\end{align}
 Here, $\Sigma_1, \Sigma_2$ are the surfaces spanned in time by the positive and negative $z-$axis, respectively. Meanwhile, our elementary nonoriented vortex, obtained from the phase $S_{\rm n}=e^{i\varphi \frac{\sigma_3}{2}}e^{i\theta \frac{\sigma_1}{2}}$, has a covariant dual field strength 
 \begin{align}
    &-\tilde{F}_{\mu\nu}= \pi \sigma_3 \int_{\Sigma} d^2\tau_{\mu\nu}\delta^{(4)}(x-y(\tau_1,\tau_2))\;, \qquad \Sigma=\Sigma_1\cup \Sigma_2\;,
\end{align}
and a dual gauge-invariant field strength $\tilde{G}_{\mu\nu}=S_{\rm n}^{-1}\tilde{F}_{\mu\nu}S_{\rm n}$:
\begin{align}
    -\tilde{G}_{\mu\nu}=\pi \sigma_3 \int_{\Sigma_1} d^2\tau_{\mu\nu}\delta^{(4)}(x-y(\tau_1,\tau_2))-\pi \sigma_3 \int_{\Sigma_2} d^2\tau_{\mu\nu}\delta^{(4)}(x-y(\tau_1,\tau_2))\;.
\end{align}
Thus,  while for a single nexus it is the {\it covariant} field strength whose orientation changes in the Lie algebra, for our nonoriented center-vortex it is the {\it gauge-invariant} field strength whose orientation changes. However, when computing the topological charge of a thin mixed configuration, the nontrivial contribution originates from the relative orientation between the terms along $S_{\rm n} \beta \cdot T S_{\rm n}^{-1} $ and $\beta \cdot T$ in Eq. \eqref{nonori-1}. Therefore, although the initial parametrizations possess rather different properties, the nonzero contribution to the total topological charge has a similar origin in both cases. \\
Now, consider $\tilde{\Sigma}$ to be a plane parallel to the $x-y$ plane, placed at a positive (negative) $z_0$ value. If this surface is completely localized at $z_0<0$ ($z_0>0$), the contributions would be given by the non-integer values $\pm 1/N$ ($\pm(N-1)/N$), where the sign depends on the plane orientation. 
This is not a surprise, since the quantization of the topological charge holds for regular gauge fields that tend to a pure gauge $A_\mu \to i U \partial_\mu U^{-1}$ when $x \to \infty$ along any direction, with single-valued $U$ in this asymptotic region. Under these conditions,
\begin{align}
     {\rm Tr}(F_{\mu\nu}\tilde{F}_{\mu\nu})|_{\rm asympt}=-\frac{2i}{3}\epsilon_{\mu\nu\rho\sigma}{\rm Tr}(\partial_\mu(U\partial_\nu U^{-1} U\partial_\rho U^{-1} U\partial_\sigma U^{-1}))\;.\label{asympt}
\end{align}
Then, following the standard arguments presented, for instance, in Ref. \cite{Rajaraman},
\begin{align}
    Q = \mathrm{const.} \int_{S_3} dS_\mu\, J_\mu \;\in \mathbb{Z} \makebox[.5in]{,}J_\mu = \epsilon_{\mu\nu\rho\sigma}{\rm Tr}(U\partial_\nu U^{-1} U\partial_\rho U^{-1} U\partial_\sigma U^{-1}) \;,\label{div-thm}
\end{align}
which measures the winding number of the map $S^3 \to U \in SU(N)$, an integer. 
In our example, the field strength extends to infinity. To obtain an integer topological charge, the field strengths should be localized in a compact region. In this respect, consider the closed and finite guiding centers $\Sigma$ and $\tilde{\Sigma}$ in Fig. \ref{mixed-thin}, where a $t=0$ time-slice is displayed. 
Instead of a nonoriented vortex on the $z$-axis, we show a finite nonoriented chain with a monopole-antimonopole pair. 
Also, instead of a plane, a spherical surface $\tilde{\Sigma}$ is considered. In the thin case, at $t \neq 0$, there is no oriented vortex, the chain maps the worldsurface $\Sigma$, and the monopole loop can be seen. At $t=0$, there are two intersection points occuring at the $\beta$ and $\beta'$ vortex branches. Therefore, as expected, the total charge is an integer $ \frac{N-1}{N} + \frac{1}{N} = +1$, and the nonoriented vortex is essential to generate a nonzero value. If the sphere $\tilde{\Sigma}$ were below the monopole, intersected at two different points of the branch with weight $\beta'$, the total charge would vanish, $Q = -\frac{1}{N} + \frac{1}{N} = 0$. That is, a nontrivial integer topological charge arises when the monopole worldline links $\tilde{\Sigma}$. This is the property obtained in the case of center vortices and nexuses~\cite{C-1,C-2}, which are likewise nonoriented but have a different parametrization from $S_{\rm n}$ in Eq.~\eqref{SW}. \\

\section {Mixed thick oriented and nonoriented center vortices}\label{mixed-thick}

\begin{figure}[h]
    \centering
    \begin{subfigure}{0.4\textwidth}
        \centering
        \includegraphics[scale=0.43]{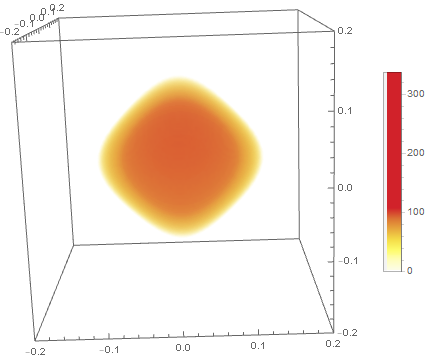}
        
        \label{fig:sub1}
        \caption{Intersecting oriented vortices with $\beta \neq \beta'$, which generate $Q=1/3$. }
    \end{subfigure}
    \hfill
    \begin{subfigure}{0.4\textwidth}
        \centering
        \includegraphics[scale=0.47]{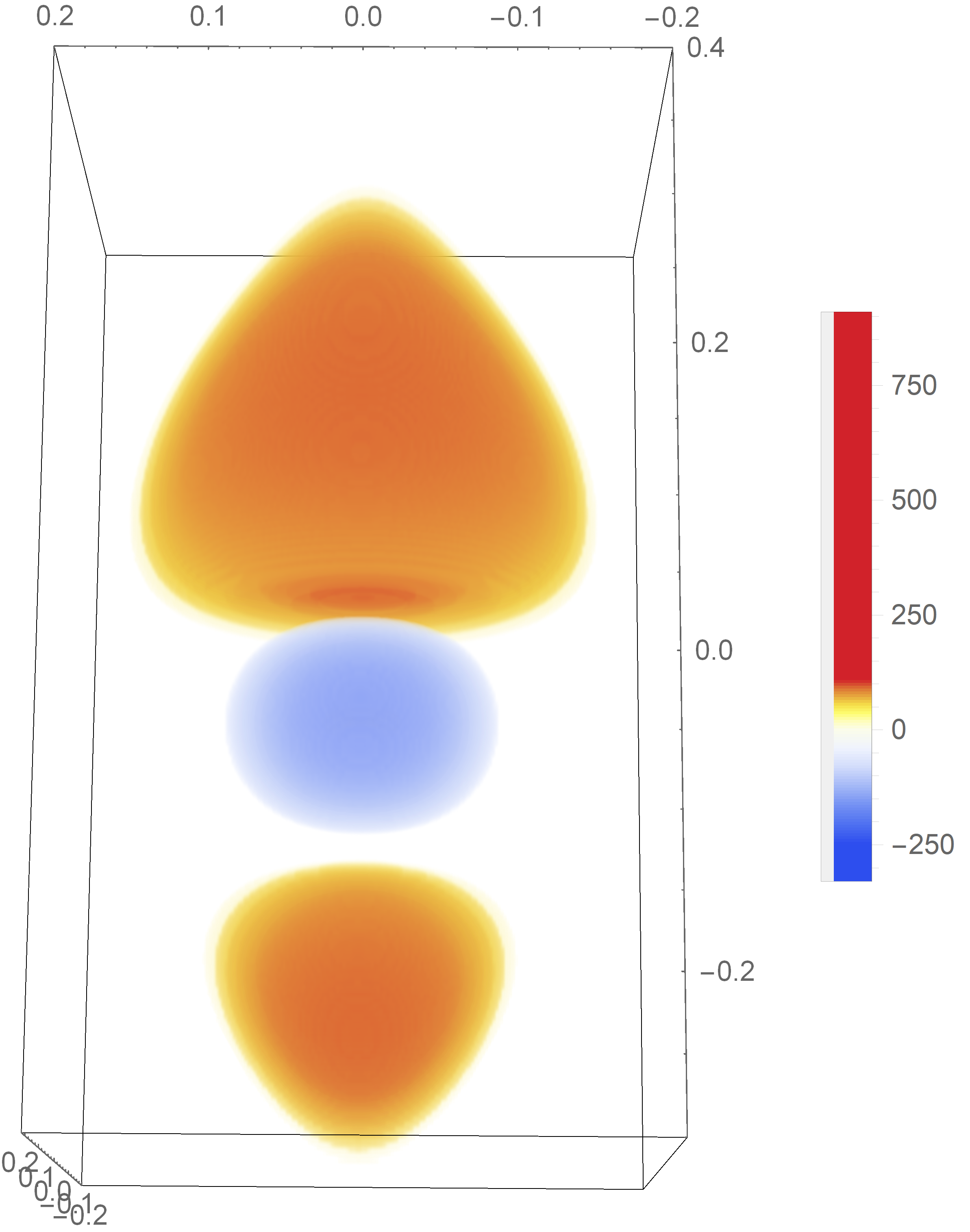}
\caption{Mixed oriented and nonoriented case}
        \label{fig:sub2}
    \end{subfigure}
    \caption{Visualization of the topological charge density $q(x)$ at $t=0$ for $SU(3)$ configurations with a thickness of $b=0.1$. The intersection of different fluxes $\beta$, $\beta'$ (a) is also realized in the lower lump of the mixed case (b), at $z\approx -0.2$. In the upper intersection around $z\approx 0.2$, the fluxes are equal, so $q(x)$ higher. A region of negative density is also formed around the monopole.}
\label{fig:ele-4}
\end{figure}

While thin center vortices are useful for gaining some understanding, they only represent the guiding centers of their physical counterparts, whose finite thickness has been measured in YM simulations \cite{vortex-th}.  
Before addressing the full non-Abelian mixed case, we first briefly extend the $SU(2)$ analysis of a pair of thick oriented planar vortices in Ref.~\cite{top-2} to $SU(N)$. For this objective, we consider guiding centers $\Sigma$ and $\tilde{\Sigma}$ 
placed at $x=0$, $y=0$ and $t=0$, $z=0$, carrying weights $\beta'$ and $\beta$, respectively. The corresponding gauge field is in the Cartan sector:
 \begin{gather}
A_\mu = a(\rho) \partial_\mu \varphi\, \beta'\cdot T + \tilde{a}(\tilde{\rho})\partial_\mu \chi\,  \beta\cdot T \;, 
\end{gather}
\begin{gather}
\rho = \sqrt{x^2 + y^2}
\makebox[.5in]{,}  \tilde{\rho} = \sqrt{z^2 + t^2}\;,
 \nonumber \\
\varphi = \arctan\!\left( \frac{y}{x} \right) \makebox[.5in]{,}
\chi = \arctan\!\left( \frac{t}{z} \right) \;,
\end{gather}  
where the scalar profiles $a$ and $\tilde{a}$ tend to zero as we approach $\Sigma$ and $\tilde{\Sigma}$, so as to control the divergence of $\partial_\mu\varphi$ and $\partial_\mu\chi$, respectively. Here, $\rho$ ($\tilde{\rho}$) is the distance to 
$\Sigma$ ($\tilde{\Sigma}$).  In addition these profiles tend to $1$ away from the cores, which map four-volumes. One of them is formed by translating a region around the $z$-axis along the Euclidean time direction.  Regarding the other core, as we move from the $t=0$ time-slice, the planar region around $z=0$ will be seen to shrink until it dissapears as $|t|$ becomes much larger than the vortex thickness. In this case, the topological charge density and total charge are
\begin{gather}
q(x) = \frac{1}{8\pi^2} \frac{\beta\cdot\beta'}{N} \frac{1}{\rho\tilde{\rho} } \frac{da}{d\rho}\frac{d\tilde{a}}{d\tilde{\rho}}  \makebox[.5in]{,}  Q=\frac{\beta\cdot\beta'}{2N}  \label{thick-or}\;.
\end{gather} 
For $SU(2)$ and $\beta' = \beta$, this yields $Q = \pm 1/2$, depending on the relative orientation of the surfaces \cite{top-2}. For $SU(3)$, the charges are $Q = \pm 1/3$ and $Q = \pm 2/3$, depending on whether $\beta' \neq \beta$ or $\beta' = \beta$, respectively. In Fig. \ref{fig:ele-4} we present a visualization of the topological charge density of the elementary intersection for $N=3$ at $t=0$. Note that with profiles $a$ and $\tilde{a}$ behaving as $\rho^2$ and $\tilde{\rho}^2$, respectively, or with higher powers, the topological charge density is regular when approaching the guiding centers.
 \\
Now, regarding the mixed oriented and nonoriented non-Abelian configuration, the expression in  Eq. \eqref{nonori} is not appropriate to implement the thickened object, as the monopole sector is not evidenced there. Our objective can be better attained by initially rewriting that gauge field in the equivalent form \cite{conf-qg}
\begin{gather}
A_\mu =   \partial_\mu \chi  P_{\beta} + \partial_\mu \varphi \left[\frac{(1 +\cos \theta)}{2}P_{{\beta}} + \frac{(1-\cos \theta)}{2} P_{{\beta}'}\right] 
- L \wedge \partial_\mu L \;,
\label{cfeten} 
\end{gather}
\begin{gather}
P_{{\beta}} = S_{\rm n}\, (\beta\cdot T) S_{\rm n}^{-1}
\makebox[.5in]{,}
P_{{\beta}'} = S_{\rm n}\, (\beta'\cdot T) S_{\rm n}^{-1}  \;.
\label{projectionb} \\
L=S_n\, \frac{(\alpha \cdot T)}{\alpha^2} S_n^{-1}= \cos \theta\, \frac{\alpha\cdot T}{\alpha^2} + \sin \theta \left( \cos \varphi 
\, \frac{T_{\bar{\alpha}}}{\sqrt{\alpha^2}}+ \sin \varphi 
\, \frac{T_{\alpha}}{\sqrt{\alpha^2}}\right) \;.
\label{elea}
\end{gather}
Eq. \eqref{cfeten} puts in evidence not only the guiding centers of the oriented and nonoriented vortices, present as singularities of $\partial_\mu \chi$ and $\partial_\mu \varphi$ at $\tilde{\Sigma}$ and $\Sigma$, respectively, but also the monopole singularity in the third term. This type of non-Abelian monopole contribution, which occurs in the asymptotic region of an 
$SU(2)$ ’t Hooft–Polyakov monopole, has been extensively studied in Refs.~\cite{Manton}–\cite{Shaba}. Indeed, the monopole appears as a correlation between the space of directions around the origin $$\hat{r}=(\cos \theta , \sin \theta \cos \varphi  , \sin \theta \sin \varphi )\;,$$ and the local Cartan direction $L$, which is an element of the local $\mathfrak{su}(2)$ subalgebra generated by
\begin{equation}
L\makebox[.5in]{,}
\frac{1}{\sqrt{\alpha^2}}\, n_{\alpha}
\makebox[.5in]{,}
\frac{1}{\sqrt{\alpha^2}}\, n_{\bar{\alpha}} \;,
\label{gener}
\end{equation}
(cf. Eq. \eqref{Tqn}). Then, we introduce the thick mixed configuration 
\begin{gather}
A_\mu =  \tilde{a} \partial_\mu \chi  P_{\beta} + a \partial_\mu \varphi \Big[\frac{(1 +\cos \theta)}{2}P_{{\beta}} + \frac{(1-\cos \theta)}{2} P_{{\beta}'}\Big] 
- h L \wedge \partial_\mu L  \;,
\label{cfeten-t}
\end{gather}
which is collimated, in spite of the presence of a monopole. Away from the monopole, where $h \to 1$, the field-strength is localized in cores around the center-vortex guiding centers $\Sigma$ and $\tilde{\Sigma}$. Outside these regions $a, \tilde{a} \sim 1$, $A_\mu$ is locally pure gauge, and $F_{\mu \nu} \sim 0$. For definitness, let us consider the simplest dependence
\begin{align}
a= a(r,\theta)\makebox[.5in]{,} a(r,0) =a(r,\pi)=0 \;, 
\label{pa}
\end{align}
where the last regularity conditions are required to produce a smooth nonoriented center vortex, localized around the $z$-axis, for every Euclidean time $t$. In addition, we shall consider an oriented vortex that is localized around a spherical surface $\tilde{\Sigma}$ of radius $r_0$ that is centered at the origin:
\begin{gather}
   \tilde{a} = \tilde{a} (r,t) \makebox[.5in]{,}  \tilde{a} (r_0,0) =0  \makebox[.5in]{,}
  \chi = \arctan\!\left( \frac{t}{r_0 - r} \right) \;.
  \label{pat}
\end{gather}
The geometry of this configuration is depicted in Fig. \ref{mixed}.

Aiming at computing the topological charge density, let us initially obtain the collimated field-strength distribution. 
Noting that $L=(P_\beta -P_{\beta'})/2$ and defining $L_0 = (P_\beta +P_{\beta'})/2$, Eq. \eqref{cfeten-t} becomes
\begin{gather}
A_\mu = A_\mu^0  + A_\mu^\alpha \makebox[.5in]{,}  A_\mu^0 = a_\mu^0  L_0  \makebox[.5in]{,} A_\mu^\alpha =  a_\mu L 
+ i h [L , \partial_\mu L] \;, \nonumber \\
a_\mu^0 = ( \tilde{a} \partial_\mu \chi  + a \partial_\mu \varphi ) \makebox[.5in]{,} a_\mu=  \tilde{a} \partial_\mu \chi + a \cos \theta \, \partial_\mu \varphi   \;.
\label{projectionb}
\end{gather}
$L_0$ involves an $S_n$-rotation of $(\beta +\beta')\cdot T$. However, $\beta+\beta'$ is minus the sum of the other $N-2$ defining magnetic weights, which satisfy $\beta_i \cdot \alpha =0$. Thus, it is simple to verify that 
\begin{gather}
 L_0 =\frac{1}{2}(\beta+\beta')\cdot T \makebox[.5in]{,} [L_0 , X]= 0 \;,
\end{gather}
where $X$ is any vector in the local $\mathfrak{su}(2)$ subalgebra generated by the elements in Eq. \eqref{gener}. 
In particular, as $A_\mu^0$ commutes with $A_\mu^\alpha$, the field strength
\begin{gather}
F_{\mu \nu} = \partial_\mu A_\nu - \partial_\nu A_\mu -i [A_\mu, A_\nu]
\end{gather}
is the sum of both field strengths computed independently:
\begin{gather}
    F_{\mu \nu} =    F_{\mu \nu}^0 +    F_{\mu \nu} ^\alpha \makebox[.5in]{,} F_{\mu \nu}^0 = f_{\mu \nu}^0 L_0 \makebox[.5in]{,} f_{\mu \nu}^0 = \partial_\mu A_\nu^0 - \partial_\nu A_\mu^0\;.
\end{gather}
The field strength for $A_\mu^\alpha$ is obtained as usual, when dealing with this type of field decomposition (see Refs. \cite{Manton}–\cite{Shaba}). As $(L,L)$ is a constant, we have:
\begin{gather}
(L, \partial_\mu L)=0 \makebox[.5in]{,}  [L, [L, \partial_\mu L]] = \partial_\mu L \makebox[.5in]{,} 
    [\partial_\mu L, \partial_\nu L] = 
    [[L, \partial_\mu L],[L, \partial_\nu L]] =X_{\mu \nu} L\;.
\end{gather}
This yields
\begin{gather}
F_{\mu\nu}^\alpha = \left( f_{\mu\nu}  +   (2h-h^2) X_{\mu \nu} \right) L  + i (\partial_\mu h) [L, \partial_\nu L] 
- i (\partial_\nu h) [L, \partial_\mu L]  + (1-h)(a_\nu \partial_\mu L - a_\mu \partial_\nu L)  \;,
\end{gather}
\begin{gather}
f_{\mu\nu} \equiv \partial_\mu a_\nu - \partial_\nu a_\mu \makebox[.5in]{,} X_{\mu \nu} = \frac{-i(L, [\partial_\mu L, \partial_\nu L])}{(L,L)} \;.
\end{gather}
Due to the orthogonality $(F_{\mu \nu}^0, F_{\rho \sigma}) = 0$, the topological charge density is obtained from
\begin{align}
q(x) = \frac{1}{32\pi^2 N} \Big ((F_{\mu\nu}^0,\tilde{F}_{\mu\nu}^0) + (F_{\mu\nu}^\alpha,\tilde{F}_{\mu\nu}^\alpha) \Big) \;.
\end{align}
After a straightforward calculation, we get
\begin{gather}
\epsilon_{\mu \nu \rho \sigma} (F_{\mu\nu}^\alpha , F_{\rho \sigma}^\alpha) 
= \epsilon_{\mu \nu \rho \sigma} \Big( f_{\mu\nu} f_{\rho\sigma}  + \xi^2 X_{\mu \nu} X_{\rho \sigma}- 2 \xi  f_{\mu\nu} X_{\rho \sigma} - 4 a_\nu \partial_\mu  \xi   X_{\rho \sigma} \Big) (L,L)  
\nonumber \\
= \epsilon_{\mu \nu \rho \sigma} \Big( f_{\mu\nu} f_{\rho\sigma}  + \xi^2 X_{\mu \nu} X_{\rho \sigma}  
- \partial_\mu[4  a_\nu   \xi   X_{\rho \sigma}  ] +4  a_\nu   \xi   \partial_\mu X_{\rho \sigma} \Big)  (L,L) \makebox[.5in]{,} \xi =  (2h-h^2)\  \;.
\end{gather}

The last term vanishes, as the derivative of $X_{\rho \sigma}$, together with the $\epsilon$-tensor, gives a contribution concentrated on the monopole worldline, where $h=0$. In addition, 
as $L$ is time-independent, there is no contribution from the quadratic term in $X$. 
Therefore,
\begin{gather}
\epsilon_{\mu \nu \rho \sigma} (F_{\mu\nu} , F_{\rho \sigma}) 
= \epsilon_{\mu \nu \rho \sigma} \Big( f_{\mu\nu}^0 f_{\rho\sigma}^0  
\Big)  (L_0,L_0) +\epsilon_{\mu \nu \rho \sigma} \big( f_{\mu\nu} f_{\rho\sigma}  
- \partial_\mu[4  a_\nu   \xi   X_{\rho \sigma}  ] \big)  (L,L)   \;.
\end{gather}
The topological charge density for the mixed configuration is displayed in Fig. \ref{fig:ele-4},
where we considered the smooth profiles, with localization length $b=0.1$,
\begin{gather}
a(\rho) = \frac{\rho^2}{\rho^2+b^2}\makebox[.5in]{,} h(r) = \frac{r^3}{r^3+b^3}\;, \nonumber \\
\tilde{a}(r,t) = \frac{\tilde{\rho}^2}{\tilde{\rho}^2+b^2}\makebox[.5in]{,}  \tilde{\rho}=\sqrt{t^2+(r-r_0)^2}  \;.
\end{gather}
 Here, $\tilde{\rho}$ measures the distance to the guiding-center $\tilde{\Sigma}$, a spherical surface with radius $r_0=0.2$ that exists only at $t=0$. As these profiles are spread and extend up to the origin, the charge is localized not only at the intersection of center-vortex branches  but also around the monopole, where a negative topological charge density is observed. The asymmetry of the densities in the upper and lower branches, around $z = +0.2$ and $z = -0.2$, arises from the different Cartan fluxes, $\beta'$ and $\beta$, carried by them. The numerical result we obtained for the total topological charge is $Q = \tfrac{1}{2}$. Once again, this fractional value does not contradict the quantization in Eq.~\eqref{div-thm}, which relies on the field being pure gauge on the three-sphere at spacetime infinity, characterized by a well-defined map $U$. These conditions are not satisfied when taking the limit to infinity while remaining inside the core of the nonoriented center vortex. 
For completeness, let us obtain the value of the total charge analytically. For the geometry proposed in Eqs. \eqref{pa} and \eqref{pat}, we have
\begin{gather}
    \epsilon_{\mu \nu \rho\sigma} f_{\mu\nu}^0 f_{\rho\sigma}^0 
  = \frac{\partial a}{\partial \theta} \Big(  \frac{\partial\tilde{a}}{\partial r} \,\partial_t \chi  -\partial_t \tilde{a} \, \frac{\partial  \chi}{\partial r} \Big)\, 8\epsilon_{ijk}  \partial_i \theta \, \partial_j \varphi \partial_k r  \;.
\end{gather}
The last factor is proportional to $1/(r^2 \sin \theta)$, which cancels with the factor of the integral in spherical coordinates. Then, the integral over $\theta$ can be performed, yielding $a(r,\pi) -a(r,0)=0$ that nulifies the contribution to $Q$. A similar calculation shows that the quadratic term in $f$ does not contribute either.

Thus, the topological charge can only be originated from 
\begin{gather}
\int d^4x\, \epsilon_{\mu \nu \rho \sigma} (F_{\mu\nu}^\alpha , F_{\rho \sigma}^\alpha) =
-4\int d^4x\,\Big(\partial_k  (a_t   \xi   X_{k}) - \partial_t  (a_k   \xi   X_{k}) \Big)(L,L) \makebox[.5in]{,}X_k=\epsilon_{ijk}X_{ij}\;.
\end{gather}
Recalling that $(L,L)= 1/\alpha^2$, and $\alpha^2 = 1/N$, we get
\begin{gather}
Q = -\frac{1}{16\pi^2}\left(\int dt \int_{S_\infty^2} ds_k \, \tilde{a} \xi \partial_t \chi X_{k} - \int d^3x\, \Delta \big(\xi a_k X_{k} \big)\right) \;,
\end{gather}
where $\Delta$ is the temporal change between $t\to +\infty $ and $t\to -\infty$: 
\begin{gather}
    \Delta \big(\xi a_k X_{k} \big)= \xi X_{k} \Delta \big( \tilde{a} \partial_k \chi \big) = \xi X_{k} \Delta \big(  \partial_k \chi \big) \;.
\end{gather}
Finally, as $\Delta \chi =-\pi$, while $\partial_k \chi$ goes to zero when $t \to \pm \infty$, we arrive at:
\begin{gather}
Q=-\frac{1}{16\pi^2 } \int dt \int_{S_\infty} ds_k \, \partial_t \chi X_{k} = - \frac{\Delta \chi}{16 \pi^2 } \int_{S_\infty^2} ds_k X_{k} = \frac{1}{2} \;.
\end{gather}
This follows from $\hat{r}$ covering the surface $S_\infty^2$ once: 
\begin{gather}
    \frac{1}{8\pi} \int_{S_\infty^2} ds_k X_k  =1\makebox[.5in]{,}  X_k=\epsilon_{ijk}\epsilon_{abc} \hat{r}_a \partial_i \hat{r}_b \partial_j \hat{r}_c \;.
\end{gather}

The total topological charge is independent of the specific smoothing profiles employed to generate the thick configurations, provided these profiles lead to regular expressions for the topological-charge density. 

\section{Mixed ensembles and the lattice}

Recently, an algorithm was proposed to identify individual topological objects in the vacuum of $SU(3)$ Yang–Mills theory \cite{frac-lump}. The authors devised a novel method to approximate the net charge of distinct topological objects within an arbitrary topological charge distribution. They are initially identified by peaks in the topological charge density and then mapped toward regions of smaller density. Measurements of the relative frequency of each type of lump were performed using two different improvement schemes for coarser and finer lattices. This made it possible to construct a histogram of their frequency for different values of the topological charge. The results showed that objects with charge $Q = 1/3$ were the most frequent.

A natural question that arises is if these observations could be explained as due to center vortices. In this regard, a wavefunctional $\Psi(A)$ peaked at an ensemble of oriented and nonoriented configurations was proposed in Ref. \cite{wavefunctional}, where $A$ is defined on a time-slice. The effective description was attained in the dual language, for Abelian-projected fields, switching from $\Psi (A)$ to the electric-field representation   $\tilde{\Psi} (E)$ via a functional Fourier transform.  In this manner, 
\begin{align}
    \tilde{\Psi}(E)=\sum_{\{\gamma\}}\psi_{\{\gamma\}}e^{i\sum\int_{\gamma}dx\cdot\Lambda_\beta}\makebox[.5in]{,}\Lambda_\beta  = 2\pi \beta|_q  \Lambda^q \;, \label{starting-pt}
\end{align}
where $\Lambda$ is a dual Cartan gauge field. 
The sum in the exponent is over the line components $\gamma$ and the corresponding distribution of weights among the $\beta_i$ within a given network $\{\gamma\}$. 
Next, we used a phenomenological probability amplitude $\psi_{\{\gamma\}}$ with phenomenological  parameters $\mu$ (tension) and $\kappa^{-1}$ (stiffness), in accordance with lattice simulations \cite{stiff1,stiff2}.
In this expression, the main building-block is the sum over all the lines carrying a given $\beta$, with fixed initial/final points and orientations, as well as fixed length $L$. An 
auxiliary scalar field $\sigma_i(x)$ coupled to the center-vortex density was also introduced to encode repulsive contact interactions between the vortex lines of the same $\beta_i$-type. These interactions were implemented through a Gaussian weight and integration over $\sigma$. 
This way, we showed that the sum over vortex networks becomes an effective field theory
 \begin{gather}
     \tilde{\Psi}(E,\eta)=\int D\Phi
     \, e^{-W} \;, \nonumber \\
     W=\int d^3x \Big(\sum_{i=1}^N\frac{1}{3\kappa}|D(\Lambda_{\beta_i})\phi_i|^2+\sum_i\big( \eta |\phi_i|^2 + \lambda |\phi_i|^4\big)-\xi\big(\phi_1 \dots \phi_N +{\rm c.c.}\big)-\vartheta \sum_{i\neq j}\bar{\phi}_i\phi_j\Big)\;.
\label{wavef}
 \end{gather} 

Similarly, based on a matrix model for noninteracting surfaces \cite{weingarten}, a partition function $Z$ that models configurations $\{\mathcal{S}\}$ formed by elementary center-vortex branches (worldsurfaces) $\mathcal{S}$, with weight distributions among the $\beta_i$,  was recently obtained in 4D Euclidean spacetime \cite{weingarten-new}. 
When no Cartan monopoles are included, it is given by
\begin{gather}
 Z_{\rm c.v.}= \frac{1}{\mathcal{N}} \int DV \exp \left( -W_{\rm c.v.}[V] \right)  \label{bp} \;, \\
    W_{\rm c.v.}[V] = -\sum_{p, i} V_i(p) + \sum_{l} \Big( \sum_i \big( \eta |V_i(l)^2|+\lambda |V_i(l)|^4 \big)-  \xi \big( V_1(l) \dots  V_N(l)+{\rm c.c.}\big)  \Big) \;,\label{abelianvortices} 
\end{gather}
where $V_i(p)$ is the plaquette variable constructed from link-variables $V_i(l) \in \mathbb{C}$, which generate branches carrying flux $\beta_i$.
To represent the complete partition function $Z$, nonoriented center vortices were included, with the monopole worldlines generated by lattice scalar fields. This plays the same role than the last term in Eq. \eqref{wavef}, which generates transitions between branches with $\beta_i \neq \beta_j$.

In both descriptions, $\xi$ controls the importance of $N$-matching, where $N$ different fluxes $\beta_i$ are matched. The parameter $\lambda$ determines the strength of the repulsive contact interaction between center vortices carrying the same Cartan weight. This is essential to stabilize the center-vortex condensate, which corresponds to $\eta < 0$. On the other hand, a direct interaction between different Cartan fluxes, $\beta_i \neq \beta_j$, is not required for this purpose. In the condensate, both approaches lead to confinement through the formation of a confining flux tube whose asymptotic string tension satisfies a Casimir law. It is clear that, due to these interactions, the model is expected to disfavor lumps arising from intersections of equal Cartan fluxes ($Q=2/3$), while the intersections involving different fluxes are not penalized. Based on the associated topological charges in these cases (cf. Eq. \eqref{thick-or}), the immediate expected consequence is that lumps with charge $Q = 1/3$ should be the most prominent.

\section{Conclusions}
\label{sec:concl}

In this work, we have presented explicit expressions for the non-Abelian gauge fields associated with thick mixed oriented and nonoriented center vortices in $SU(N)$ Yang–Mills theory. The non-Abelian phases ensured a continuous interpolation between different Cartan fluxes at monopole junctions. By introducing suitable profile functions, the gauge field was rendered smooth everywhere.  We analyzed and visualized the resulting topological charge density, whose lava-lump morphology reflects the underlying geometry and the properties of the Cartan subalgebra of $\mathfrak{su}(N)$. 

In a future work, it would be interesting to compute the topological susceptibility implied by the wavefunctional peaked at center-vortex guiding centers proposed in Ref.~\cite{wavefunctional}, where the dependence on the vortex thickness is expected to play a central role. A comparison of typical topological charge-density distributions with lattice data would also be valuable, as it could provide deeper insight into the nonperturbative structure of the Yang–Mills vacuum.
In this direction, we emphasize that lumps arising from the intersection of center-vortex branches carrying different elementary Cartan fluxes $\beta, \beta'$ in $SU(3)$ have total charge $Q=1/3$, while intersections of equal elementary fluxes yield $Q=2/3$. Notably, in Ref. \cite{frac-lump}, the charge of lumps detected in $SU(3)$ lattice Yang–Mills simulations was found to be $Q=1/3$. When comparing with the center-vortex framework, at a first glance, the number of possibilities combining different Cartan fluxes is expected to be larger than for equal fluxes. Furthermore, the wave functional \cite{wavefunctional} and the surface model \cite{weingarten-new} for Abelian-projected configurations naturally incorporate noninteracting and repulsively interacting branches for different and equal fluxes, respectively. In this way, the vortex condensate becomes stabilized while the main asymptotic properties of confining flux tubes are reproduced. It is noteworthy that this mechanism favors intersections between branches carrying different weights, and therefore the emergence of $Q=1/3$ lumps, as observed in the lattice.

\section{Acknowledgments}
This work was partially
supported by Funda\c{c}\~ao de Amparo \`a Pesquisa do Estado de S\~ao Paulo (FAPESP), grants nos. 2023/18483-0 (D. R. J.), 2024/20896-3  (D. R. J.) and 2023/11867-7 (G. M. S.). Financial support from the Brazilian agency CNPq under Contract No. 309971/2021-7 (L. E. O.) is also
gratefully acknowledged.

\end{document}